\documentclass{emulateapj}

\shorttitle{Quasar Mid-Infrared Luminosity Evolution}
\shortauthors{Singal et al.}

\slugcomment{to appear in ApJ}

\begin{document}
\title{THE MID-INFRARED LUMINOSITY EVOLUTION AND LUMINOSITY FUNCTION \\ OF QUASARS WITH {\it WISE} AND SDSS}

\author{J. Singal, J. George, A. Gerber}

\affil{Physics Department, University of Richmond\\28 Westhampton Way, Richmond, VA 23173, USA}

\email{jsingal@richmond.edu}

\begin{abstract}
We determine the 22$\mu$m luminosity evolution and luminosity function for quasars from a data set of over 20,000 objects obtained by combining flux-limited Sloan Digital Sky Survey optical and Wide field Infrared Survey Explorer mid-infrared data.  We apply methods developed in previous works to access the intrinsic population distributions non-parametrically, taking into account the truncations and correlations inherent in the data.  We find that the population of quasars exhibits positive luminosity evolution with redshift in the mid-infrared, but with considerably less mid-infrared evolution than in the optical or radio bands.  With the luminosity evolutions accounted for, we determine the density evolution and local mid-infrared luminosity function.  The latter displays a sharp flattening at local luminosities below $\sim 10^{31}$\,erg\, s$^{-1}$\,Hz$^{-1}$, which has been reported previously at 15 $\mu$m for AGN classified as both type-1 and type-2.  We calculate the integrated total emission from quasars at 22 $\mu$m and find it to be a small fraction of both the cosmic infrared background light and the integrated emission from all sources at this wavelength.
\end{abstract}

\section{Introduction} \label{intro}

As different processes in active galaxies (AGN) result in emission in different energy ranges, a crucial class of information for understanding black hole, accretion disk, and jet systems is the intrinsic population characteristics of AGN in widely separated wavebands \citep[e.g.][]{Dermer07,La10}.  These tell us how AGN have evolved in different wavebands over the history of the Universe, as well as their luminosity and spectral distributions, and the correlations among the different waveband emissions.  Knowledge of these are necessary to constrain models of emission mechanisms as well as establish luminosity functions and integrated outputs at given wavelengths and their relation to those at other wavelengths.  This work focuses on the mid-infrared population properties of quasars, an important class of AGN, using a dataset of over 20,000 quasars seen with the Wide Field Infrared Explorer satellite \citep[{\it WISE} ---][]{WISE}.  As we wish to focus on the mid-infrared wavelengths that are most distinct in emission from the near-infrared, we are interested in particular in the properties of quasars in the longest {\it WISE} wavelength 22 $\mu$m band.

Multiple strategies are possible for compiling a dataset of AGN from an infrared survey to determine infrared population properties.  On the one hand, one can select AGN candidates using color or spectral-based selection in the infrared.  This technique has been utilized with combined {\it Spitzer} Infrared Space Telescope and {\it WISE} data by \citet{Lacy} who determine luminosity functions at 5 $\mu$m.  Alternately, one can select potential AGN candidates from infrared colors and perform optical followups and/or matches to optical catalogs to classify objects and derive redshifts.  This technique has been used with combined Infrared Space Observatory and Infrared Astronomical Satellite data by \citet{Matute06} who, using spectral fits, determine 15 $\mu$m luminosity functions.  Alternately, one could use optically identified quasars from the overlap of a deep optical survey area and a deep infrared survey.  This has been utilized with {\it Spitzer} data by \citet{Brand06} and \citet{Babbedge06} who present luminosity functions at 8 $\mu$m and 3.6, 4.5, 5.8, 8, and 24 $\mu$m, respectively.  Of these techniques the first two have the advantage of detecting AGN which are obscured in the optical \citep[see e.g.][]{Mateos12,Mateos13} but the potential disadvantages of missing AGN which do not display the requisite infrared colors, and the inability to classify AGN as specifically quasars or another type.  The third technique has the advantage of being complete in the optical for a class of AGN but could miss other optically obscured objects.  Some of the aforementioned works have the advantage of not being substantially flux-limited in the optical, either because of the use of infrared selection criteria for identification or the use of deep fields or both.

However, in contrast to their advantages, all three of the previously listed compilation techniques have the disadvantage of having a small number of AGN objects, numbering in the hundreds, or using photometric redshifts with complicated incompletness and selection considerations in the case of \citet{Babbedge06}.  They also may be subject to complex or incompletely understood selection effects in one or more wavebands such as in \citet{Matute06}, and require spectral modeling in many cases.  

In this work we are interested in performing a complementary determination of AGN population characteristics in the mid-infrared with a dataset covering a large portion of the sky with objects numbering in the tens of thousands, with complete spectroscopic redshifts, in which the selection effects are known and in which the crucial population characteristics are determined directly from the data non-parametrically with minimal modeling and assumptions.  This can be done by restricting to quasars and using {\it WISE} data, along with techniques which we have developed.  

Given that the largest catalogs of quasars with redshifts are identified and catalogued by their optical spectral characteristics, a large dataset such as this used to evaluate population characteristics of quasars in another band depends on an optical survey as well and thus the limits of that survey.  In order to evaluate the luminosity evolution in both mid-infrared and optical, and to separate and compare these effects, we require a dataset that has both infrared and optical fluxes to reasonable and known limits across a broad range of redshifts.  The overlap of the {\it WISE} satellite AllWISE catalog with the Sloan Digital Sky Survey (SDSS) quasar catalog \citep{SDSSQ}, can form such a dataset.

When dealing with data from a large survey in the $a$ waveband, the luminosity function is usually obtained from a flux limited sample $f_{j,a} > f_{j_m,a}$ with $f_{j_m,a}$ denoting the flux limit of the $j$th object and the luminosity being $L_{j,a} = 4 \, \pi \, d_L^2 f_a / K_a$, where $d_L(z)$ is the luminosity distance and $K_a(z)$ stands for the K-correction.  For a pure power law emission spectrum of index $\varepsilon_a$ defined as $f_a \propto \nu^{-\varepsilon_a}$, one has $K_a(z) = (1+z)^{1-\varepsilon_a}$.  This simple form may be augmented by the presence of emission lines, as in the optical data in this work.  

In general, the determination of the full luminosity function and its evolution requires analysis of the bi-variate distribution $\Psi_a(L_a,z)$.  A correlation between $L_a$ and $z$ is known as luminosity evolution and would need to be taken into account when determining the distributions of the individual variables $L_a$ and $z$.  In the case of quasars here with the optical and some other band luminosity, because an optical measurement is necessary for quasar identification and spectroscopic redshift, we have at least a tri-variate function. We must take into account not only the correlations between the redshift and individual luminosities (i.e. the two luminosity evolutions) but also the possible intrinsic correlation between the two luminosities, before individual distributions can be determined \citep[e.g.][]{QP2}.  Treating the infrared survey data as a stand-alone sample independent of the optical survey truncations and the relations between infrared and optical luminosities is not appropriate.

\citet{EP92,EP99} pioneered new methods for determining the correlation of variables from a flux limited and more generally truncated dataset, which were expanded to multiwavelength and complicatedly truncated cases in works by Singal et al. \citep[e.g.][]{QP1,QP2,BP2}.  Our aim in this paper is to take all the selection and correlation effects into account in determination of the true evolution of optical and mid-infrared luminosities of quasars and to find their distributions, using an SDSS $\times$ {\it WISE} dataset.   

In \S \ref{datasec} we describe the infrared and optical data used.  \S \ref{simlumf} contains a general discussion of luminosity evolution and the sequence of the analysis.  In \S \ref{evsec} we apply the methods to achieve the luminosity-redshift evolutions and the correlation between the luminosities.  We determine the density evolution in \S \ref{dev}, and the local luminosity functions in \S \ref{local}.  In \S \ref{testass} we investigate some of the assumptions used and their effect on uncertainty, and \S \ref{disc} contains a discussion of the results.  This work uses the standard $\Lambda$CDM cosmology with $H_0=71$\,km\,s$^{-1}$\,Mpc$^{-1}$, $\Omega_{\Lambda}=0.7$ and $\Omega_{m}=0.3$.

\section{Data} \label{datasec}

\begin{figure}
\includegraphics[width=3.5in]{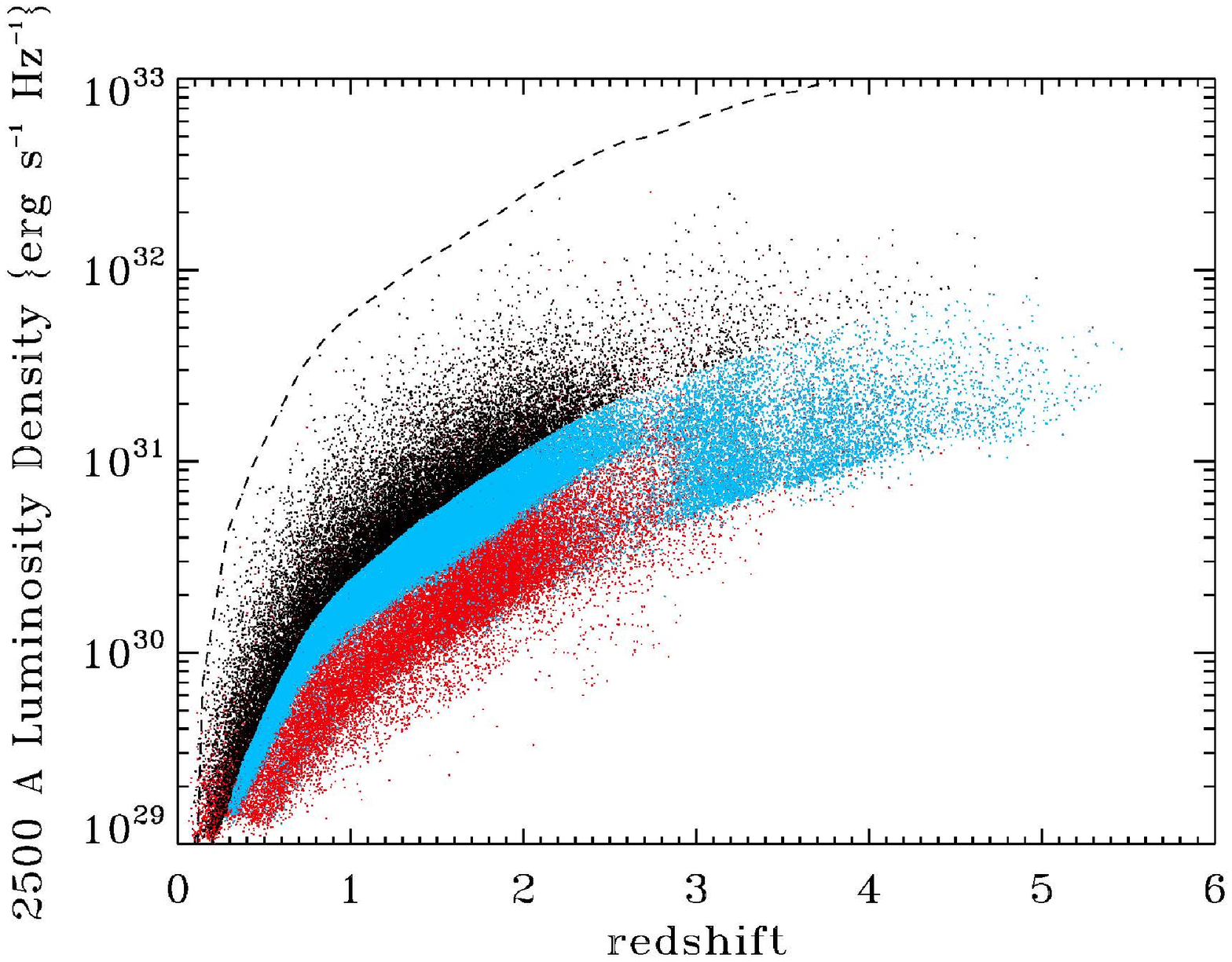}
\caption{The 2500 \AA \, rest frame absolute luminosity density for all 105,760 objects in the SDSS DR7 quasar catalog \citep{SDSSQ}.  The 2500 \AA \, luminosity density is determined from the observed $i$-band magnitude as described in \S \ref{datasec} and includes applying the K-corrections provided by \citet{R06} which include the continuum and emission line effects.  Objects plotted in red are those who do not have target flags indicating that they were flagged for spectroscopic followup based on optical colors or a radio match, or are flagged as extended sources, and are not considered for this analysis (see \S \ref{datasec}).  Of the remaining 62,276 objects, those plotted in blue are those which do not meet the K-corrected $i<19.1$ criterion (see \S \ref{datasec}), while black points are the objects that do meet the K-corrected $i<19.1$ criterion (21,600 objects) and are used as the parent optical population in this analysis.  It is seen that the K-corrected $i<19.1$ and flagged subset forms a catalog that has a somewhat smoother redshift distribution with a reduced a bias toward objects with $z>2$ (although still with residual biases in the redshift distribution as discussed further in \S \ref{dev}), and with a calculable limiting flux for every redshift.  The solid line is the upper limiting flux corresponding to $i=$15.0.  }
\label{opts}
\end{figure} 

\begin{figure}
\includegraphics[width=3.5in]{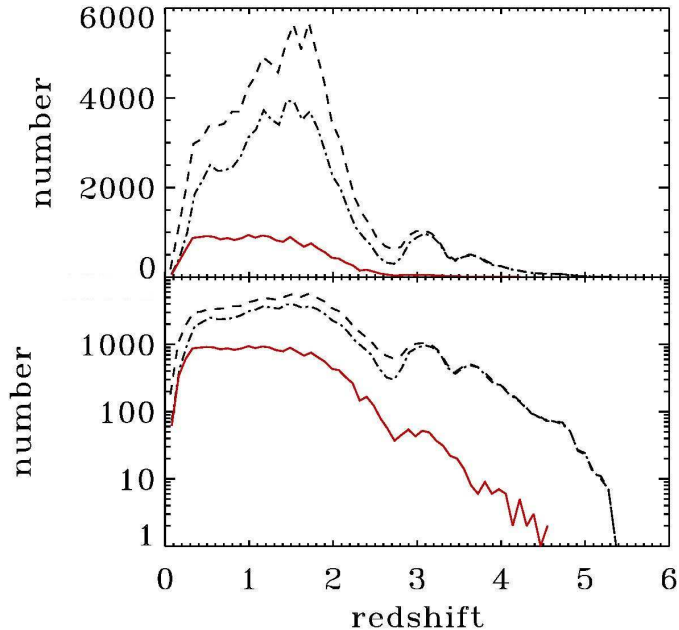}
\caption{Histogram of redshifts of quasars in three different sample cuts in bins redshift 0.1 wide, displayed in linear (top) and logarithmic (bottom) scale.  The dashed line represents the full 105,760 objects in the SDSS DR7 quasar catalog.  The dash-dot line represents those selected for spectroscopic followup based on either optical colors or a radio match (51,190 objects), and the solid line represents the ones selected for followup based on these criteria and which meet the K-corrected $i<19.1$ criterion (21,600 objects) which is used as the parent optical sample for this analysis.  The effect of the non-uniform selection function with redshift is discussed in \S \ref{dev}.}
\label{optsreds}
\end{figure} 

\begin{figure}
\includegraphics[width=3.5in]{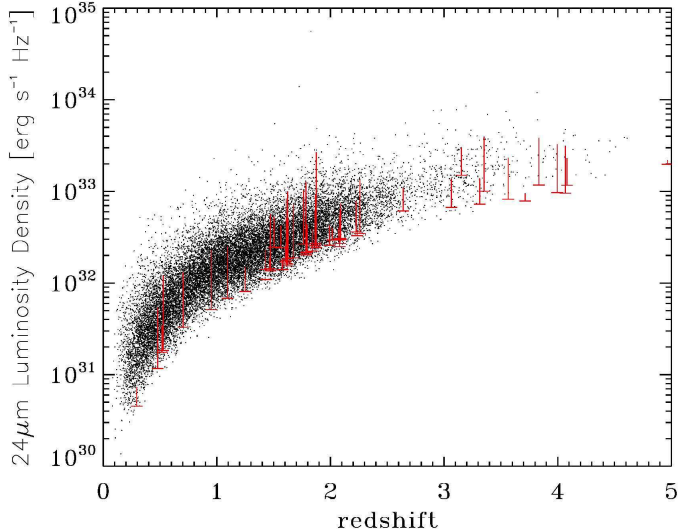}
\caption{The 22 $\mu$m rest frame absolute luminosity density for the quasars in the SDSS $\times$ {\it WISE} sample used in this analysis (20,063 objects).  To obtain the 22 $\mu$m GHz luminosity density we use the luminosity distance obtained from the redshift and the standard cosmology and the K-correction discussed in \S \ref{datasec}.  For a few objects (but only a few flor clarity) the bottom of the red line indicates the lower limit luminosity for inclusion of the particular object in the survey.}
\label{influms}
\end{figure} 

\begin{figure}
\includegraphics[width=3.5in]{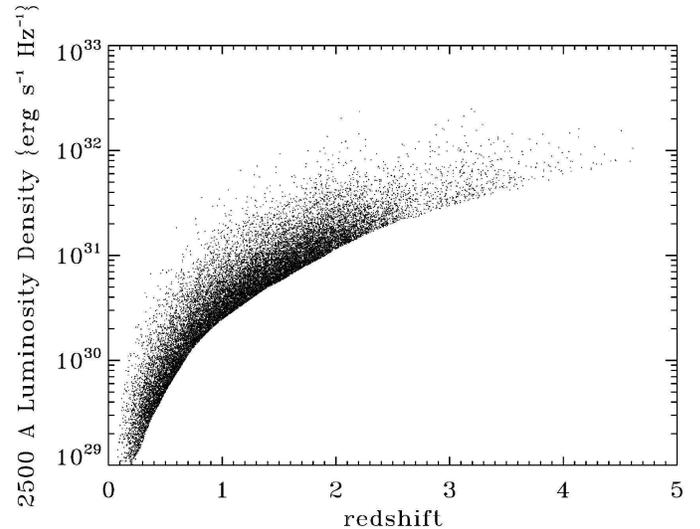}
\caption{Similar to figure \ref{opts} but showing only the quasars in the main SDSS $\times$ {\it WISE} sample used in this analysis (20,063 objects). }
\label{optlums}
\end{figure}

We use the well-established SDSS data release 7 (DR7) quasar catalog, which contains over 105,000 objects.  We seek a subset with a well-defined flux-limit for inclusion for every object and a reduced bias in the redshift distribution.  In the SDSS DR7, objects were identified as quasar candidates for spectroscopic follow-up if they displayed the requisite optical colors, or if identified via ROSAT X-ray data, or if selected by a  so-called ``serendipity'' algorithm that identifies unusual colors in concert with a radio match, or if they had a radio match within $2''$ \citep{SDSSQ}.  This means that the full DR7 quasar catalog cannot be considered to be truly optically flux-limited in at any given magnitude limit.  

To obtain a set that is flux-limited in the optical band, we restrict the full set in two ways.  First we allow only those objects that display the requisite optical colors for follow-up or are identified via a radio match within $2''$ (the latter criterion applies to less than 1 \% of objects), and only those objects which are unextended.  This corresponds to requiring that the target flag be set to ``1'' in one of three columns in the DR7 quasar catalog --- the Low-z target selection flag (\#55), the High-z target selection flag (\#56), and the FIRST selection flag (\#57) --- and that a flag be set to ``0'' for extended morphology (column \#33), and reduces the full catalog from 105,760 to 62,276 objects.  Next we impose a K-corrected $i$-band magnitude limit of 19.1 at redshift 2.  As discussed in \citet{SDSSQ} and \citet{R06} this results in a catalog with a smoother redshift distribution with a reduced bias toward objects with $z>$2 (although still with residual biases in the redshift distribution as discussed further in \S \ref{dev}).  It results in a catalog that is flux-limited at every redshift, although that flux limit is somewhat redshift dependent.  The magnitude criterion reduces the set further to 21,600 objects.  We consider this as the parent set of identified quasars for which we seek an infrared match. The luminosities of the DR7 quasars are shown in Figure \ref{opts}, and the raw redshift distributions of subsamples discussed here are shown in Figure \ref{optsreds}.  We also note the presence of an upper limit $i$ band magnitude of 15.0 for inclusion in the catalog.  However this criterion is not completely rigorous, as mentioned in \citet{SDSSQ}, and corresponds to a luminosity higher than almost all sources. 

The AllWISE catalog is an extension to the {\it WISE} general all-sky data release that combines data from the cryogenic and post-cryogenic \citep[NEOWISE --- ][]{Mainzer11} survey phases.  It contains over 700 million objects observed by {\it WISE} in the 3.4, 4.6, 12 and 22 $\mu$m mid-infrared bandpasses (known as W1, W2, W3, and W4 respectively).  We match the AllWISE catalog with the restricted set of SDSS DR7 quasars with a two arcsecond matching radius criterion, resulting in 20,063 matches with a detection in the 22$\mu$m band, defined as having a signal-to-noise ratio (SNR) value greater than two.  There are zero multiple matches with this matching radius.  Reducing the matching radius to one arcsecond results in only a slight reduction of 3\% in the number of matches.  Therefore two arcseconds is an appropriate choice to include real matches while eliminating spurious ones.  Over 95\% of the parent flux-limited optical set have an infrared match with a detection in the 22$\mu$m band.

For the K-corrections to calculate luminosities in the optical band, we adopt the full $\varepsilon_{\rm opt}=0.5$ power law continuum {\it plus} emission line K-corrections presented in \citet{R06} and discussed in \S 5 and Table 4 of that work.  The methods of this work can then account for any bias resulting from emission line effects, as long as they are included in the conversions from luminosity to flux, i.e. in the K-corrections.  For the 22 $\mu$m infrared K-correction, we note that studies have generally shown that quasar spectral energy distributions transition from being roughly flat ($\varepsilon \sim 0$) in L$_{\nu}$ space between 100 and $\sim$ 10 $\mu$m to having $\varepsilon \sim 1$ at shorter wavelengths \citep[e.g.][]{Gallagher07,Richards06b}.  In light of this we adopt a 24$\mu$m K-correction scheme in which the K-correction has $\varepsilon = 0$ [$K(z)$ = 1+$z$] for $z <$ 1.4 and $\varepsilon = 1$ [$K(z)$ = 2.4] for $z >$ 1.4 with a smooth leveling off.  We adopt this population-averaged K-correction scheme in part to avoid the additional truncation and bias complications that would result from requiring an observation of every object in the sample in every other {\it WISE} band in order to fit an infrared spectrum.  Figures \ref{influms} and \ref{optlums} show the infrared and optical luminosities versus redshifts of the quasars in the constructed SDSS x {\it WISE} sample.  

The limiting optical flux for inclusion in the data set is a function of redshift, and corresponds to the $i$-band magnitude 19.1 at $z$=2 K-corrected to the redshift in question.  It thus has a slight dependence on redshift, as can be seen in Figure \ref{optlums}.  On the other hand, the limiting 22 $\mu$m infrared flux for inclusion in the survey, and therefore the limiting 22 $\mu$m luminosity for inclusion of a particular object, depends on that object's SNR in that waveband.  In particular, since an SNR of 2 is the lowest SNR for an object to have a flux reported at 22 $\mu$m, the lower limiting flux in that band (W4) for an object $j$ to be included in the sample is
\begin{equation}
F_{j,W4,lim} = F_{j,W4} \times { {2} \over {SNR_j}}
\end{equation}
The minimum limiting 22 $\mu$m luminosity that an object could have to be included in the survey is a function of its lower limiting flux and redshift, with the standard luminosity-redshift relation
\begin{equation}
L_{j,22 \mu m,lim}(z) = {{ 4 \pi \, D_L\!(z) \, {F_{j,W4,lim} } } \over {K\!(z)} }
\label{lumlimeq}
\end{equation}
and $z=z_j$.  Some minimum limiting 22 $\mu$m luminosities are shown in Figure \ref{influms}.

We have also explored using subsequent data releases of SDSS quasars from the Baryon Oscillation Spectroscopic Survey (BOSS) phase of the project \citep{SDSSQ12}.  As the BOSS project is optimized to detect galaxies in a certain redshift range and not to produce a complete flux-limited quasar catalog while minimizing detection biases, we find that these catalogs are inferior to the DR7 catalog for our purposes.  The BOSS catalog actually results in fewer matches to {\it WISE} sources than the DR7.  

We have also performed the analysis in this work with a significantly less restricted parent optical set, consisting of the DR7 quasars that simply have an $i$-band magnitude less than 19.1 and neglecting the target flags and the K-corrections to the magnitude for inclusion.  This results in a uniform flux optical limit for inclusion, with a parent optical set of 63,492 objects and a matched 22 $\mu$ and optical set of 49,415 objects.  The major conclusions for the luminosity evolutions, the correlations between the luminosities, and the local luminosity functions obtained using that alternate data set are quite similar to those obtained with the more restrictive data set, indicating a robustness of the results in this work.

\section{General remarks on luminosity functions and evolutions}\label{simlumf}

\subsection{Luminosity and density evolution}\label{lde}

The luminosity function gives the number of objects per unit comoving volume $V$ per unit source luminosity, so that the number density is $dN/dV = \int dL_{\rm a} \Psi_{\rm a}(L_a, z)$ and the total number is $N = \int dL_{\rm a} \, \int dz \, (dV/dz) \, \Psi_{\rm a}(L_a, z)$.  To examine luminosity evolution, without loss of generality, we can write a luminosity function in some waveband $a$ as 

\begin{equation}
\Psi_{a}\!(L_a,z) = \rho(z)\,\psi_a\!(L_{a}/g_{a}\!(z) , \eta_a^j)/g_a\!(z),
\label{lumeq}
\end{equation}
where $g_{a}\!(z)$ and $\rho(z)$ describe the luminosity evolution and comoving density evolution with redshift respectively and $\eta_a^j$ stands for parameters that describe the shape (e.g. power law indices and break values) of the $a$ band luminosity funtion.  In what follows we assume a non-evolving shape for the luminosity function (i.e. $\eta_a^j= const$, independent of $L$ and $z$), which is a good approximation for determining the global evolutions.  The later point is discussed in \S \ref{testass}.  Once the luminosity evolution $g_{a}\!(z)$ is determined using the methods described below we can obtain the mono-variate distributions of the independent variables $L'_{\rm a}=L_{\rm a}/g_{\rm a}(z)$ and $z$, namely the density evolution $\rho(z)$ and ``local'' luminosity function $\psi_{\rm a}$.  The total number of observed objects is then
\begin{equation}
N_{tot} = \int_0^{z_{max}} dz \int_{L_{\rm min}(z)}^\infty { dL_{a} \, \rho(z) \, {dV \over dz} \, { {\psi_a\!\left(L_{a}/g_{a}\!(z)\right)} \over {g_a\!(z)} }  } ,
\label{inteq}
\end{equation}

We consider this form of the luminosity function for luminosities in different bands, allowing for separate (optical and infrared) luminosity evolutions.  We use a parameterization for the luminosity evolution with redshift
\begin{equation}
g_{\rm a}(z)={ {(1+z)^{k_{\rm a}}} \over {1+({{1+z} \over {z_{cr}}})^{k_{\rm a}}} }.
\label{altevolution}
\end{equation} 
which has been shown to be a goot fit for a dataset based on SDSS with many $z>3$ objects \citep{QP2}.  As discussed in that work, good value of $z_{cr}$ is 3.7, but the precise value does not matter for the analysis.  With $g_a(0) \approx 1$ for positive values of $k_a$ the luminosities $L'_a$ refer to the de-evolved values at $z=0$, hence the name ``local''. 

We discuss the determination of the evolution factors $g_a(z)$, which in this parameterization becomes a determination of $k_a$, in \S \ref{evsec}.  The density evolution function $\rho(z)$ is determined by the method shown \S \ref{dev}.  Once these are determined we construct the local (de-evolved) luminosity function $\psi_{\rm a}\!(L_{a}')$ as in \S \ref{local}.

\subsection{Joint Luminosity Functions}\label{ifcorr}

In general, determination of the evolution of the luminosity function of extragalactic sources with spectroscopic redshifts for any wavelength band except optical involves a tri-variate distribution because spectroscopic and most photometric redshift determination requires optical observations which introduces additional observational selection bias and data truncation.  Thus, in a case such as this, unless redshifts are known for all sources in an infrared survey from infrared data alone, we need to determine the combined luminosity function $\Psi\!(L_{\rm opt},L_{\rm IR},z)$ from a tri-variate distribution of $z$ and the fluxes in the optical and mid-infrared bands.  If the optical and mid-infrared luminosities were statistically independent variables, then this luminosity function would be separable in the form of $\Psi\!(L_{\rm opt},L_{\rm IR},z) = \Psi_{\rm opt}(L_{\rm opt}, z) \, \times \, \Psi_{\rm IR}(L_{\rm IR}, z)$ and we would be dealing with two bi-variate distributions.  

However, there may also be a correlation between the two luminosities.  As described below, the methods employed here allow us to determine whether any pair of variables are independent or correlated.  If it is determined that the luminosities are correlated (see \S \ref{roevsec}), the question must be asked how much of this luminosity correlation is intrinsic to the population and how much is induced in the data by flux limits and/or similar luminosity evolutions with redshift.  Determination of this is quite intricate as discussed in e.g. \citet{PS15} and Appendix B of \citet{QP1}, and has not been explored sufficiently in the literature.  While this will be the subject of future investigations, here we will consider both possibilities.  

At one extreme, if the luminosity correlation is intrinsic and not induced, one should seek a coordinate transformation to define a new pair of variables which are independent.  This requires a functional form for the transformation.  We define a new luminosity which we call a ``correlation reduced infrared luminosity'' $L_{\rm cri}=L_{\rm IR} / F(L_{\rm opt} / L_{\rm fid})$, where the function $F$ describes the correlation between $L_{\rm IR}$ and $L_{\rm opt}$ and $L_{\rm fid}$ is a fiducial luminosity taken here\footnote{This is a convenient choice for $L_{\rm fid}$ as it is lower than the lowest 2500 \AA \, luminosity considered in our sample, but results do not depend on the particular choice of numerical value.} to be $10^{28}$\,erg\, s$^{-1}$\,Hz$^{-1}$.  For the correlation function we will assume a simple power law 
\begin{equation}
L_{\rm cri} = {{L_{\rm IR}} \over {(L_{\rm opt}/L_{\rm fid})^{\alpha}}}
\label{rcrdef}
\end{equation}
where $\alpha$ is a bulk power law correlation index to be determined from the data.  This is essentially a coordinate rotation in the log-log luminosity space.  As shown in \S \ref{evsec} below, we can determine a best fit value for the index $\alpha$ which orthogonalizes the new luminosities.  Given the correlation function we can then transform the data (and its truncation) into the new independent pair of luminosities $(L_{\rm opt}$ and $L_{\rm cri})$.  The local luminosity functions of uncorrelated luminosities $L'_{\rm opt}$ and $L'_{\rm cri}$ can then be used to recover the local infrared luminosity function by a straight forward integration over $L'_{\rm cri}$ and the true local optical luminosity function as 
\begin{eqnarray}
\psi_{\rm IR}\!(L_{\rm IR}') = 
\nonumber \\ \int_0^{\infty} { \psi_{\rm opt}\!(L_{\rm opt}') \, \psi_{\rm cri}\left({{ L_{\rm IR}' } \over ({{L_{\rm opt}'/L_{\rm fid}})^{\alpha} } } \right) \, {{dL_{\rm opt}'} \over ({{L_{\rm opt}'/L_{\rm fid}})^{\alpha}}  } \,}
\label{localrad}
\end{eqnarray}
The mid-infrared luminosities also undergo luminosity evolution with 
\begin{equation}
g_{\rm IR}\!(z) = g_{\rm cri}\!(z) \, \times \, [g_{\rm opt}\!(z)]^{\alpha}
\label{gradform}
\end{equation}
(cf equation \ref{rcrdef}).

At the other extreme, if the correlation between the luminosities is entirely induced by truncation effects and similar redshift evolutions, then the luminosity functions are separable into $\Psi_{\rm opt}(L_{\rm opt}, z) \, \times \, \Psi_{\rm IR}(L_{\rm IR}, z)$ as described above and the analysis can proceed from there.   

As noted above, we will consider both possibilities here as extreme cases.  It turns out that the major results obtained in both cases are very similar.

\section{Determination of best fit correlations} \label{evsec}

Here we first give a brief summary of the algorithmic strategy involved in these determinations, which was first proposed by Efron and Petrosian and has been expanded upon in recent works.  This method uses a modified rank test to determine the best-fit values of parameters describing the correlation functions using the test statistic 

\begin{equation}
\tau = {{\sum_{j}{(\mathcal{R}_j-\mathcal{E}_j)}} \over {\sqrt{\sum_j{\mathcal{V}_j}}}}
\label{tauen}
\end{equation}
to test the independence of two variables in a dataset, say ($x_j,y_j$) for  $j=1, \dots, n$.  Here $\mathcal{R}_j$ is the dependent variable ($y$) rank of the data point $j$ in a set associated with it, $\mathcal{E}_j=(1/2)(n+1)$ is the expectation value and $\mathcal{V}_j=(1/12)(n^{2}+1)$ is the variance, where $n$ is the number of objects in object $j$'s associated set.  For untruncated data (i.e. data truncated parallel to the axes) the set associated with point $j$ includes all of the points with a lower (or higher, but not both) independent variable value ($x_k < x_j$).  If the data is truncated one must form the {\it associated set} consisting only of those points of lower (or higher, but not both) independent variable ($x$) value {\it that would have been observed if they were at the $x$ value of point $j$ given the truncation} (see e.g. \citet{BP2} for a full discussion of these points). 

If ($x_j,y_j$) are independent then the ranks $\mathcal{R}_j$ should be distributed randomly and $\tau$ should sum to near zero.  Independence is rejected at the $m \, \sigma$ level if $\vert \, \tau \, \vert > m$.  To find the best fit correlation bewteen $y$ and $x$ the $y$ data are adjusted by defining $y'_j=y_j/F(x_j)$  and the rank test is repeated, with different values of parameters of the function $F$ until $y'$ and $x$ are determined to be uncorrelated.

In this analysis we can ignore the upper optical flux limit of SDSS quasars discussed in \S \ref{datasec}.  The reason for this is that data truncations are only consequential in this analysis if the truncation is actually depriving the sample of data points that exist.  As can be seen in Figure \ref{opts}, there are very few objects approaching the optical upper truncation limit, indicating that this truncation does not appreciably alter the sample from the underlying population.  All of the truncations, therefore, are one-sided --- i.e. at the lower end of fluxes and luminosities.

\subsection{\rm Infrared-Optical Luminosity Correlation} \label{roevsec}

\begin{figure}
\includegraphics[width=3.5in]{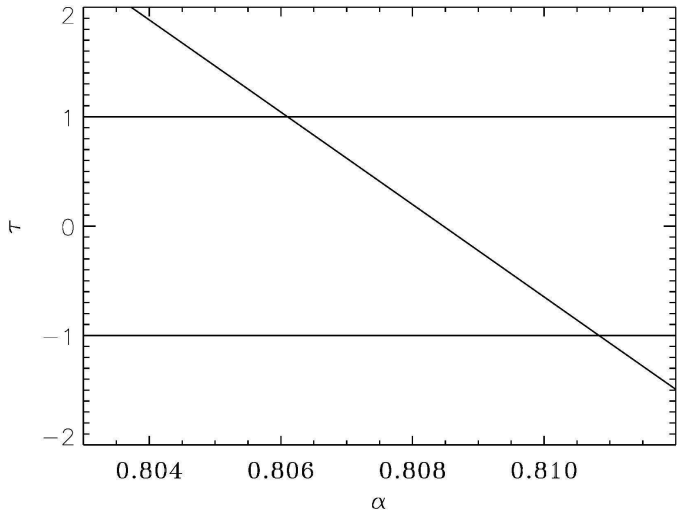}
\caption{The value of the $\tau$ statistic as given by equation \ref{tauen} as a function of $\alpha$ for the observed relation $L_{\rm IR} \propto (L_{\rm opt})^{\alpha}$, where $L_{\rm opt}$ and $L_{\rm IR}$ are the optical and infrared luminosities, respectively, for the quasars in the combined sample.  The 1$\sigma$ range for the best fit value of $\alpha$ is where $\vert \, \tau \, \vert \leq 1$.  It is seen that the observed optical and infrared luminosities are positively correlated, but with a sub-linear relation, although this may not be the true intrinsic correlation in this case, as discussed in \S \ref{roevsec}. } 
\label{taus}
\end{figure}

As an example of the determination of a correlation, here we will determine the observed correlation between the observed infrared and optical luminosities.  Assuming the correlation function between the luminosities of the form of equation \ref{rcrdef} we calculate the test statistic $\tau$ from equation \ref{tauen} as a function of $\alpha$.  Where $\tau$ is closest to zero corresponds to the values of $\alpha$ that remove the correlation.  Figure \ref{taus} shows the absolute value of the test statistic $\tau$ vs $\alpha$, from which we get the best fit value of $\alpha=0.8185$ with one $\sigma$ range $\pm 0.005$.  

As discussed in \S \ref{ifcorr}, this correlation may be inherent in quasars or may be a result of the data truncations and similar positive redshift evolutions.  The general yet quite nuanced and often overlooked question of determining whether an observed correlation between different waveband luminosities is intrinsic or induced will be explored in a future work.  To complete the present analysis in the most robust manner, we will consider both possibilities.  We shall see that it does not make a significant difference for the major conclusions of this work.  It is interesting to note that the observed power-law correlation for the infrared and optical luminosities seen here is less than that for radio and optical luminosities ($\alpha \sim 1$) seen in \citet{QP1} and \citet{QP2}.

\subsection{Joint Dataset Luminosity-Redshift Correlations} \label{method}

The basic method for determining simultaneously the best fit intrinsic $k_{\rm opt}$ and $k_{\rm IR}$, given the evolution forms in equation \ref{altevolution} and properly taking into account the data truncations, is more complicated because we now are dealing with a three dimensional distribution  ($L_{\rm IR}, L_{\rm opt}, z$) and two correlation functions ($g_{\rm IR}\!(z)$ and $g_{\rm opt}\!(z)$), plus we can find the true intrinsic correlation in this case because the truncation effects in the luminosity-redshift space are known and redshift is the independent variable in both cases.

Since we have two criteria for truncation, the associated set for each object $k$ includes only those objects that are sufficiently luminous in both bands to have been in the survey if they were located at the redshift of the object in question.  As discussed in \S \ref{datasec}, for the optical data, this would be all objects with a luminosity greater than the limiting optical luminosity at the redshift of object $k$ given the optical flux limit as a function of redshift, while for the 22 $\mu$m infrared data this is all objects whose luminosity is greater than their minimum limiting luminosity calculated at the redshift of object $k$ (i.e. equation \ref{lumlimeq} with $z=z_k$).

The luminosity cutoff limits for a given redshift must also be adjusted by factors of $g_{\rm opt}\!(z)$ and $g_{\rm IR}\!(z)$. Consequently, we have a two dimensional minimization problem, because objects will drop in and out of associated sets as $g_{\rm opt}\!(z)$ and $g_{\rm IR}\!(z)$ change, leading to changes in the calculated ranks in equation \ref{tauen}.

\begin{figure}
\includegraphics[width=3.5in]{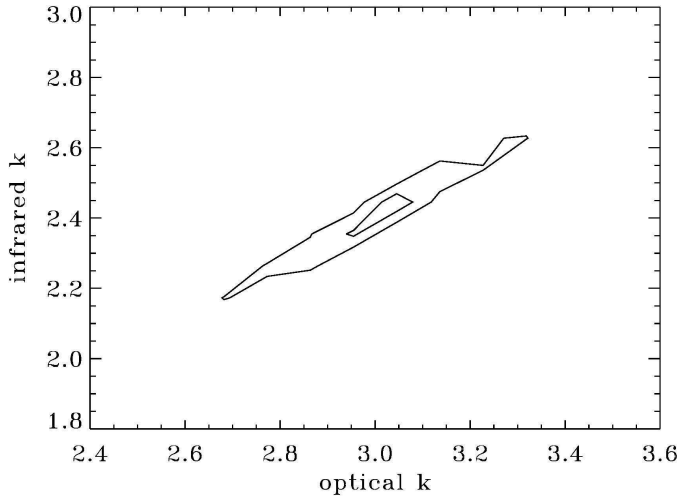}
\caption{The $1\sigma$ and $2 \sigma$ contours for the simultaneous best fit values of $k_{\rm opt}$ and $k_{\rm IR}$ of the combined infrared-optical sample, for the forms of the luminosity evolutions given by equation \ref{altevolution}.   }
\label{alphas}
\end{figure}

We form a test statistic $\tau_{\rm comb} = \sqrt{\tau_{\rm opt}^2 + \tau_{\rm IR}^2} $ where $\tau_{\rm opt}$ and $\tau_{\rm IR}$ are those evaluated considering the objects' optical and mid-infrared luminosities, respectively.  The favored values of $k_{\rm opt}$ and $k_{\rm IR}$ are those that simultaneously give the lowest $\tau_{\rm comb}$ and, again, we take the $1 \sigma$ limits as those in which $\tau_{\rm comb} \, < 1$.  Figure \ref{alphas} shows the 1 and 2 $\sigma$ contours for $\tau_{\rm comb}$ as a function of $k_{\rm opt}$ and $k_{\rm IR}$.

We see that positive evolution in both infrared and optical wavebands is favored.  The minimum value of $\tau_{\rm comb}$ favors an optical evolution with $k_{\rm opt}$ = 3.0 $\pm$ 0.1 and an infrared evolution with $k_{\rm IR}$ = 2.4 $\pm$ 0.1.  It should be noted that  $k_{\rm opt}$ as determined here from the combined infrared-optical dataset is quite similar to that determined from both the a much larger optical only dataset with only a $i$-band magnitude cut (3.3 $\pm$ 0.1) and a combined optical-radio dataset (3.0 --- 3.5 depending on the radio flux limit assumed) in \cite{QP2}, indicating that the truncations have been properly handled and the robustness of the method.  We have previously verified this method including with monte carlo simulations as discussed in e.g. \citet{QP1} and \citet{QP2}.

\begin{figure}
\includegraphics[width=3.5in]{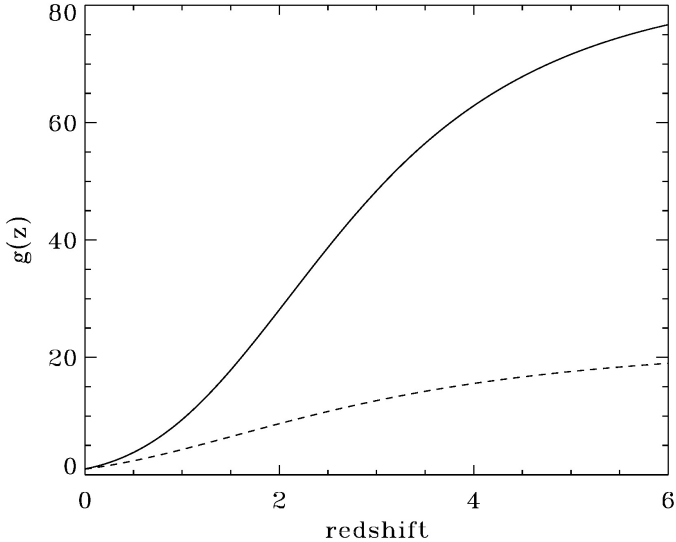}
\caption{Plots of $g_{\rm opt}(z)$ (solid) and $g_{\rm IR}(z)$ (dashed) vs. $z$ for the functional form of equation \ref{altevolution} and with the middle best-fit $k_{\rm opt}$ and $k_{\rm IR}$ values determined in this analysis. } 
\label{gs}
\end{figure}

If we consider that the infrared-optical luminosity correlation is entirely inherent in the underlying data, then the orthogonal luminosities are $L_{\rm opt}$ and $L_{\rm cri}$ (see \S \ref{ifcorr}) and we can determine the best fit evolutions $g_{\rm cri}\!(z)$ and $g_{\rm opt}\!(z)$.  These results favor $k_{\rm opt}=3.6 \pm 0.2$ and $k_{\rm cri}=-0.4 \pm 0.2$.   In this case the best fit infrared evolution can be recovered by equation \ref{gradform} and it would have $k_{\rm IR}=2.4$ at low redshifts and a more complicated form at higher redshifts, which corresponds well with the results obtained from considering $k_{\rm opt}$ and $k_{\rm IR}$ as orthogonal.  The optcal evolutions found by the two methods are in tension at the 1$\sigma$ level although within 2$\sigma$ agreement.  For visualization in Figure \ref{gs} we plot the functions $g_{\rm opt}(z)$ and $g_{\rm IR}(z)$ vs. $z$ for the middle of the best-fit $k_{\rm opt}$ and $k_{\rm IR}$ values determined by this analysis.

These results indicate that quasars have undergone a significantly lesser evolution in mid-infrared luminosity relative to optical luminosity, and indeed relative to radio luminosity where $k_{\rm rad}$ = 5.5 \citep{QP2}.  We return to this point in \S \ref{disc}.

For comparison in the literature, \citet{Babbedge06} state that optically identified quasars have an mid-infrared luminosity evolution, when fit to the form $(1+z)^\gamma$, of $\gamma \sim 3$, while \citet{Matute06} find an exponent with that functional form of 2.9 for luminosities at 15 $\mu$m, both of which are somewhat stronger evolution than the result here for redshifts up to $\sim 3$, at which point the functional forms diverge from the one employed here at higher redshifts.  

\section{Density evolution} \label{dev}

Next we determine the density evolution $\rho(z)$.  One can define the cumulative density function 

\begin{equation}
\sigma(z) = \int_0^z { {{dV} \over {dz}} \, \rho\!(z) \, dz}
\end{equation}
which, following \citet{P92} based on the method of \citet{L-B71} which is equivalent to a maximum likelihood estimate, can be calculated by

\begin{equation}
\sigma(z) = \prod_{j}{(1 + {1 \over m(j)})}
\label{sigmaeqn}
\end{equation}
where the set of $j$ includes all objects with a redshift lower than or equal to $z$, and $m(j)$ is the number of objects with a redshift lower than the redshift of the object at redshift $z$ {\it which are in that object's associated set}.  In this case, the associated set is again those objects with sufficient optical and radio luminosity that they would be seen if they were at redshift $z$.  The use of only the associated set for each object accounts for the biases introduced by the data truncation.  Then the density evolution $\rho(z)$ is 

\begin{equation}
\rho\!(z) = {d \sigma\!(z) \over dz} \times {1 \over dV/dz}
\label{rhoeqn}
\end{equation}

However, to determine the density evolution, the luminosity evolution determined in \S \ref{roevsec} must be taken out.  Thus, the objects' optical and infrared luminosities, as well as the optical and infrared luminosity limits for inclusion in the associated set for given redshifts are scaled by taking out factors of $g_{\rm opt}\!(z)$ and $g_{\rm IR}\!(z)$ which are determined as above.  The preceding method is fully adequate if there is a uniform selection function across redshift for quasars at a given flux.  A non-uniform selection function will bias the redshift distribution by artificially removing objects at certain redshifts.  As shown in e.g. Figure 6 of \citet{R06} the selection function of SDSS quasars is not uniform across redshift.  As stated in that work, by restricting to sources that are not extended, have been selected for spectroscopic followup based on either optical colors or a radio match, and are brighter than $i$=19.1 magnitude after applying the emission line K-correction, the redshift distribution can be made smoother but not unbiased.  Indeed this is visible in Figure \ref{optsreds} of this work.  To correct for the bias due to the non-uniform selection function, we adopt the selection function derived in \citet{R06} and correct the derived redshift distribution at each redshift for the incompleteness.  In particular, if at the redshift and magnitude of any given object $j$ the survey is only a fraction $m$ complete then the differential density function at that redshift $\rho(z_j)$ should be increased by a factor of 1/$m$, and the cumulative density function $\sigma(z_j)$ at that redshift should be increased by a factor of  1/$m$ $\times$ ($\sigma(z_j)$ - $\sigma(z_{j-1})$).  We apply these corrections, including a ``floor'' on the selection function of 0.333 as implemented in \citet{R06}, to obtain the intrinsic density evolution functions $\sigma(z)$ and $\rho(z)$. The extent of the selection function correction can be seen in Figure \ref{rholog}. 

\begin{figure}
\includegraphics[width=3.5in]{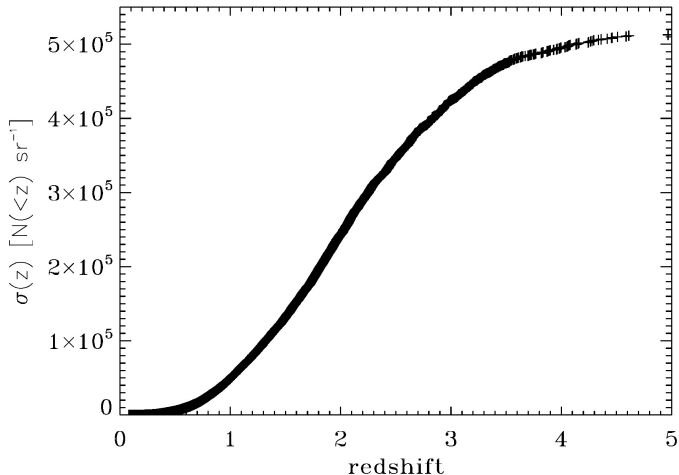}
\caption{The cumulative density function $\sigma(z)$ vs. redshift for the quasars calculated from the sample in this work.  The normalization of $\sigma(z)$ is determined as described in \S \ref{dev}.  A spline fit to $\sigma\!(z)$ is used to determine $\rho\!(z)$ by equation \ref{rhoeqn}.  }
\label{sigma}
\end{figure}

\begin{figure}
\includegraphics[width=3.5in]{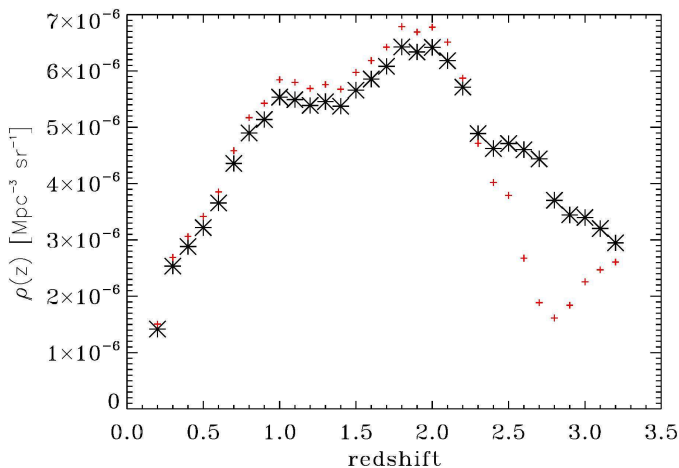}
\caption{The density evolution $\rho(z)$ vs. redshift (large stars) for the for the quasars, calculated from the sample in this work.  $\rho(z)$ is defined such that $\sigma(z)=\int_0^{\infty} \rho(z) \, dV/dz \, dz$.  The normalization of $\rho(z)$ is determined as described in \S \ref{dev}. For reference we also overplot the form of $\rho(z)$ if the selection function correction discussed in \S \ref{dev} is not applied (small red crosses).  The offset between the two determinations at redshifts below $z=2.3$ is primarily because of the change in normalization of $\rho(z)$ determined by equation \ref{inteq}. }
\label{rholog}
\end{figure}

Figures \ref{sigma} and \ref{rholog} show the cumulative and differential density evolutions, respectively. The normalization of $\rho(z)$ is determined by equation \ref{inteq}, with the customary choice of $\int_{L_{\rm min}'}^{\infty} {\psi(L') \, dL'}=1$.  The number density of quasars seems to peak at just below a redshift of 2.  This is later than found in \citet{MP99} and \citet{QP2} but similar to the results in \citet{Shaver96}, and \citet{Hopkins07} and slightly earlier than the results in \citet{R06}.   We note that we plot $\rho(z)$ only to redshift 3.2 because the number of objects is rapidly falling at that redshift and the distribution is more prone to errors resulting from small fluctuations in numbers.  We also note that $\rho(z)$ as plotted here contains a factor of dV/dz as in equation \ref{rhoeqn}.

We note that the analysis of this work in principle accesses the intrinsic redshift distribution for the full range of local luminosities ($L_{\rm opt}'$, $L_{\rm IR}'$) and redshift present in the data set, accounting for the various truncations.  Outside of the range of local luminosities and redshifts present in the data set the population could differ in some systematic way from within the range present in the data set.  The data set is effectively unlimited in maximum luminosity, and contains local luminosities down to $5 \times 10^{28}$\,erg\, s$^{-1}$\,Hz$^{-1}$ in optical and $3 \times 10^{30}$\,erg\, s$^{-1}$\,Hz$^{-1}$ in mid-infrared, so outside of this range the redshift distribution could differ from the one obtained here.  The result obtained here for the redshift distribution is the intrinsic bulk average for the full range of luminosities present in the data set.  An analysis of if and how the redshift distribution differs for various ranges of mid-infrared and/or optical luminosities would require considering a different form than equation \ref{lumeq} with $\eta_a^j= const$ for the full luminosity function in a band and is beyond the scope of this work, a point which we return to in \S \ref{testass}.

\section{Local luminosity functions} \label{local}

\subsection{General Considerations}

In a parallel procedure to determine the redshift distribution we can use the local (redshift evolution taken out, or 'de-evolved') luminosity (and de-evolved luminosity thresholds) to determine  the `local' luminosity distributions, i.e. the luminosity functions $\psi_{a}\!({L_a}')$, where again the subscript $a$ denotes the waveband, and the prime indicates that the luminosity evolution has been taken out.  We first obtain a cumulative luminosity function

\begin{equation}
\Phi_a\!(L_{a}') = \int_{L_{a}'}^{\infty} {\psi_a\!(L_{a}'') \, dL_{a}''}
\end{equation}
which, following \citet{P92} using the method of \citet{L-B71}, $\Phi_a\!(L_{a}')$, can be calculated by

\begin{equation}
\Phi_a\!(L_{a}') = \prod_{k}{(1 + {1 \over n(k)})}
\label{phieq}
\end{equation}
where $k$ runs over all objects with a luminosity greater than or equal to $L_a$, and $n(k)$ is the number of objects with a luminosity higher than the luminosity of object $k$ which are in object $k$'s associated set, which in this case consists of those objects which would be in the survey if they were at object $k$'s luminosity considering the luminosity limits for inclusion in both optical and mid-infrared. 
The luminosity function $\psi_a\!(L_{a}')$ is 

\begin{equation}
\psi_a\!(L_{a}') = - {d \Phi_a\!(L_{a}') \over dL_{a}'}
\label{psieqn}
\end{equation}

In \S \ref{evsec} we determined the luminosity evolutions for the optical luminosity $L_{\rm opt}$ and the mid-infrared luminosity $L_{\rm IR}$.  We can form the local optical $\psi_{\rm opt}\!(L_{\rm opt}')$ and mid-infrared $\psi_{\rm IR}\!(L_{\rm IR}')$ luminosity functions straightforwardly, by taking the evolutions out.  As before, the objects' luminosities, as well as the luminosity limits for inclusion in the associated set for given redshifts, are scaled by taking out factors of $g_{\rm IR}\!(z)$ and $g_{\rm opt}\!(z)$, with $k_{\rm IR}$ and $k_{\rm opt}$ determined in \S \ref{evsec}.  We use the notation $L \rightarrow L' \equiv L/g(z) $.  For the local luminosity functions, we use the customary normalization $\int_{L_{\rm min}'}^{\infty} {\psi(L') \, dL'}=1$.  This normalization may be biased by around 8\% due to quasar variability as discussed in \S \ref{testass}.

\subsection{Local optical luminosity function}\label{locopt}

\begin{figure} 
\includegraphics[width=3.5in]{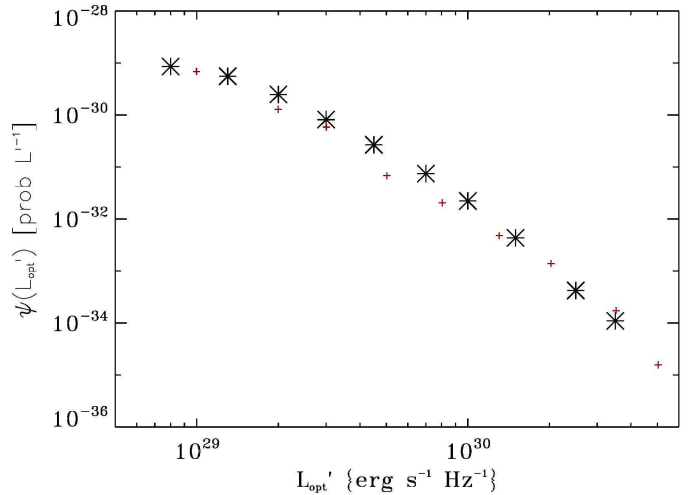}
\caption{The local optical luminosity function $\psi_{\rm opt}\!(L_{\rm opt}')$.  The large stars show the results from the sample in this work while the small red crosses show the results from a combined optical-radio dataset from \citet{QP2}.   The normalization of the local luminosity functions is described in \S \ref{local}.  The power law slopes of $\psi_{\rm opt}\!(L_{\rm opt}')$ are discussed in \S \ref{locopt}. }
\label{psiopt}
\end{figure}

Figure \ref{psiopt} shows the local differential $\psi_{\rm opt}\!(L_{\rm opt}')$ optical luminosity of the quasars calculated from the sample, and for comparison that determined for a combined optical-radio sample in \citet{QP2}.  We would expect these to be very similar since presumably the underlying population is the same although subjected to different selection effects in the different samples.

The optical luminosity function shows possible evidence of a break at $\sim 10^{30}$\,erg\, s$^{-1}$\,Hz$^{-1}$. Fitting a broken power law above and below $\sim 10^{30}$\,erg\, s$^{-1}$\,Hz$^{-1}$ yields values for the power law slopes of $-2.3 \pm 0.2$ and $-4.3 \pm 0.1$ below and above the break, respectively.  For comparison, these values were found to be $-2.8\pm 0.3$ and $-3.8\pm 0.5$ for the combined optical-radio dataset, and $-2.8\pm 0.2$ and $-4.1\pm 0.4$ for the parent optical-only SDSS dataset, in \citet{QP2}.  As mentioned in that work, the optical luminosity function has been studied extensively in various AGN surveys.  For example, \citet{Boyle00}, using the 2dF optical dataset (but with no radio overlap criteria) use a customary broken power law form for the luminosity function, with values ranging from $-$1.39 to $-$3.95 for different realizations, showing reasonable agreement.  We note that as discussed in \S \ref{testass} the normalization of the local optical luminosity function may be biased by as much as 12 percent.

\subsection{Local mid-infrared luminosity function}\label{localinf}

\begin{figure}
\includegraphics[width=3.5in]{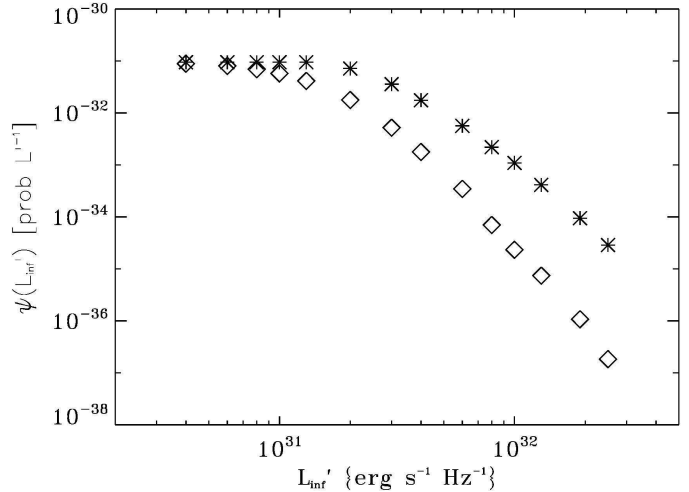}
\caption{The local 22 $\mu$m mid-infrared luminosity function $\psi_{\rm IR}\!(L_{\rm IR}')$ for quasars, calculated from the combined infrared-optical sample.  The diamonds show the results from considering that the infrared-optical luminosity correlation is entirely induced, while the stars show the results from considering the correlation to be entirely intrinsic, calculated with equation \ref{localrad}.  We consider the range between these extremal cases to be the overwhelming source of undertainty in the determination of $\psi_{\rm IR}\!(L_{\rm IR}')$. The power law slopes of $\psi_{\rm IR}\!(L_{\rm IR}')$ and the flattening at local mid-infrared luminosities below $2 \times 10^{31}$\,erg\, s$^{-1}$\,Hz$^{-1}$ are disussed in \S \ref{localinf}. }
\label{psiinf}
\end{figure}

Figure \ref{psiinf} shows the local 22 $\mu$m mid-infrared luminosity function $\psi_{\rm IR}\!(L_{\rm IR}')$ calculated both the extremal cases of a) assuming that the optical and infrared luminosities are truly independent (stars) and b) assuming that all of the observed correlation in the luminosities is intrinsic and constructing the local infrared luminosity function from $\psi_{\rm opt}\!(L_{\rm opt}')$ and $\psi_{\rm cri}\!(L_{\rm cri}')$  with equation \ref{localrad} (diamonds).  The two determinations vary somewhat and we take this to be the overwhelmingly dominant source of uncertainty in the reconstructed intrinsic local infrared luminosity function.

It is seen that this local mid-infrared luminosity function contains a strong break around $\sim 2 \times 10^{31}$\,erg\, s$^{-1}$\,Hz$^{-1}$ with a dramatic flattening at luminosities below the break.  The flattening is also seen by \citet{Matute06} at 15 $\mu$m for both type-1 and type-2 AGN at approximately the same value of luminosity as seen here (assuming a relatively flat spectrum from 15 $\mu$m to 22 $\mu$m), and at 24 $\mu$m for AGN in the heavily model-dependent analysis of \citet{Xu01}.  We discussed some implications of this flattening in \S \ref{disc}.  We note that a differential luminosity function which is flat at the faint end corresponds to a cumulative luminosity function which has a power law slope of -1 at the faint end, i.e. that the number of objects still increases with decreasing luminosity but not dramatically.

For luminosities above the break we determine a power law slope of $-3.9 \pm 0.9$, with this relatively large range resulting from the difference in considering the luminosity correlation intrinsic versus induced.  For comparison \citet{Matute06} report values for the bright end power law slope at 15 $\mu$m ranging between -2.13 and -3.15 depending on model assumptions. \citet{Babbedge06} plot a local 24 $\mu$m luminosity function for luminosities above $\sim 10^{31}$\,erg\, s$^{-1}$\,Hz$^{-1}$ that is similar to the overplotted \citet{Matute06} luminosity function in that range.

\section{Tests of assumptions and error considerations}\label{testass}

A few considerations point to the robustness of the results obtained here and previously.  As discussed in \S \ref{method}, results for the optical and infrared evolution factors determined here are consistent whether the infrared and optical luminosities are considered to be truly orthogonal or not.  Also as mentioned there the best-fit optical evolution factor $k_{\rm opt}$ determined with the combined infrared-optical sample here is quite similar to that determined previously with both the parent optical only sample and a combined radio-optical sample.  As discussed in \S \ref{locopt}, the local mid-infrared luminosity function shows the same dramatic flattening at low luminosities whether the infrared and optical luminosities are considered to be truly orthogonal or not.  Likewise the local optical luminosity function $\psi_{\rm opt}\!(L_{\rm opt}')$ as determined with the combined infrared-optical sample here is quite similar to that determined previously with both the parent optical only sample and a combined radio-optical sample.  Reults here for the mid-infrared luminosity function parallel those in other works, as do the power laws for the redshift evolution of the infrared luminosity to the extent that results are comparable.

Additionally, as mentioned in \S \ref{datasec}, we have also performed the analysis in this work with a significantly less restricted parent optical set, consisting of the DR7 quasars that simply have an $i$-band magnitude less than 19.1 and neglecting the target flags and the K-corrections to the magnitude for inclusion, resulting in a much parent larger set with a uniform flux optical limit for inclusion.  The major conclusions for the luminosity evolutions, the correlations between the luminosities, and the local luminosity functions using that alternate data set are quite similar to those obtained with the more restrictive data set, indicating a robustness of the results.  

We also emphasize that even if the truncations that we apply on the parent optical quasar catalog to create a flux-limited data set (including if flux limits vary by redshift) need not correspond exactly to the actual flux limit of the survey, as long as the limit is consistently applied in both creating the catalog and performing the analysis --- this point is addressed in \citet{BP1}.

We consider here some possible sources of error.

{\it Luminosity dependent density evolution}:  A possible concern is that luminosity dependent density evolution, which is not explicitly considered in the functional forms for the luminosity functions used here (i.e. equation \ref{lumeq} with $\eta_a^j= const$), may be necessary to represent the evolution of the luminosity function.  As a test of whether the functional forms used here are inadequate for the considered analysis, we divide the data into high and low sets of de-evolved luminosity $L'$ (cutting on optical luminosity at a middle value of $3 \times 10^{29}$\,erg\, s$^{-1}$\,Hz$^{-1}$) and check the similarity of the computed density evolutions for the two sets and for the whole combined set.  The density evolutions computed for both cuts are similar to each other, with the both the high half and the low half peaking in $\rho(z)$ at $z=1.8$.  Given the similarity of these distributions to each other and to that computed from the dataset as a whole, we conclude that we are justified in neglecting an explicit luminosity dependent density evolution form for the purposes of this analysis, which depends on a full luminosity function of the form of equation \ref{lumeq}.  While luminosity dependent evolution is a well-fit model explored in many works, the form of equation \ref{lumeq} with $\eta_a^j= const$ is also adequate for the ranges of (local) luminosities and redshifts considered in this analysis, in particular for recovering in the bulk average intrinsic redshift distribution for objects with luminosities in the full range present in the data set.  As mentioned in \S \ref{dev} an analysis of if and how the redshift distribution differs for different combinations of mid-infrared and optical luminosities would require considering a different form than equation \ref{lumeq} with $\eta_a^j= const$ for the full luminosity function in a band and is beyond the scope of this work.

{\it Survey measurement errors}:  It is well known that measurement errors in magnitudes have the potential to bias the results if the number density of sources increases with decreasing flux, since it is more likely that sources will be erroneously included than excluded \citep{Eddington40}.  The extent of this effect depends on the faint end source counts power law slope, and in the limit of flat differential source counts there is no bias.  In the limit of measurement errors which are constant in fractional flux, an error will be introduced on the {\it normalization} of the source counts, and therefore on that of the luminosity functions, and can be approximated by [$1/2 \, \delta^2 \, m_{below} \, (m_{below}+1)$] where $\delta$ is the fractional error in flux and $m_{below}$ is the faint end differential source counts power law slope \citep{Teer04}.   Although we have obtained the intrinsic local luminosity functions and not the source counts for quasars explicitly, we observe that in the sample the lowest fluxes correspond to roughly the lowest decade of local luminosities for both bands, so will approximate the faint end of the source counts power law slope with that of the faint end of the local luminosity functions to consider this effect.  For the mid-infrared luminosity function $m_{below} \sim 0$ so we would expect no appreciable error on the normalization of the mid-infrared luminosity function from this effect.  For the optical luminosity function $m_{below} \sim 2.7$, and the typical reported SDSS measurement errors are on the order of a few hundredths of a magnitude \citep{SDSSQ}  but we will conservatively adopt errors of 0.2 magnitudes in $i$ band to account for both measurement errors and the intrinsic RMS scatter due to source variability.  For the faint end magnitudes of 19.1 and this error, $\delta$ is around 0.16, so the bias on the optical luminosity function {\it normalization} will be 12 percent or less.  

On the other hand, there will be an effect on the reconstructed power law {\it slope} of the luminosity functions only if the fractional measurement errors change systematically with luminosity.  We can pursue an upper limit on this effect by considering a related quantity which is readily available from the data sets --- how fractional measurement errors in flux depend on flux.  The optical data show only a modest dependence on $i$ band reported error with magnitude at magnitudes below 19.1, with magnitude errors at most a factor of 1.5 higher at the highest magnitude end of that range than for the lowest magnitudes, which corresponds to a small differential fractional flux error.  The 22 $\mu$m data show a stronger dependence of the reported error in the measured magnitude on the magnitude leading to reported fractional errors in flux which are three to four times higher for the faintest fluxes as for the brightest.\footnote{We note though that the majority of objects do not have a reported quantified error on the W4 magnitude.}  However data at a given flux corresponds to a wide range of luminosities, especially at the lowest fluxes.  For example as mentioned above in both bands the objects with the lowest fluxes densely span an order of magnitude in local luminosity, which would significantly wash out the systematic dependence of fractional errors on flux when considering if fractional errors in luminosity vary systematically with luminosity.  Thus we consider that the dependence of fractional errors in luminosity on luminosity are small enough in both bands, which along with the above consideration that the faint end differential source counts are relatively flat in the 22 $\mu$m band, leads to the conclusion that any effect on the power law slopes of either of the luminosity functions is negligible.  

{\it Redshift bias in the SDSS quasar catalog}:  One could ask if the density evolution determined here is affected by any biases toward certain redshifts in the SDSS quasar sample.  As stated in \citet{R06}, the main sources of bias in the redshift distribution of the SDSS quasar sample are 1) the differing magnitude limits for $z>2$, 2) the effects of emission lines on $i$ band flux at different redshifts, and, at a somewhat less important level, 3) the inclusion of extended sources at the lowest redshifts ($z \sim 0.3$).  As discussed in \S \ref{datasec}, we have dealt with issues 1 and to some extent issue 2 by restricting the sample to a universal K-corrected $i<19.1$ magnitude limit and by adopting for calculating luminosities the \citet{R06} K-corrections which include the effect of emission lines as well as the continuum spectrum, and issue 3 by not including extended sources.  We further address issue 2 by incorporating the SDSS quasar selection function derived in \citet{R06} and correct the derived redshift distribution at each redshift for the incompleteness as discussed in \S \ref{dev}.  We do not believe that the density evolution determined is significantly affected by biases between the redshifts of 0 and 3 where the objects are overwhelmingly most common.  There may be biases above redshift 3 but they do not affect the major conclusions of this work.

\section{Discussion}\label{disc}

We have used a general and robust method to determine the mid-infrared and optical luminosity evolutions and luminosity functions simultaneously for quasars using a SDSS $\times$ {\it WISE} dataset, which combines 22 $\mu$m infrared and $i$-band optical data for over 20,000 quasars ranging in redshifts from 0.08 to 4.97.

As discussed in \S \ref{intro} quite different strategies can be used to assemble an infrared AGN data sample for determination of infrared population characteristics.  These strategies have advantages and disadvantages.  Here we have chosen to assemble a large sample of tens of thousands of objects with definite spectroscopic redshifts and known and straightforward flux truncations for inclusion, from which the true intrinsic population characteristics of interest can be determined directly and non-parametrically with limited modeling and assumptions.  

\subsection{Luminosity Evolutions}

Here we find, as discussed in \S \ref{method}, that quasars have undergone significant luminosity evolution with redshift in the mid-infrared, but less than in the optical band, and, comparing to previous results \citep[e.g.][]{QP1,QP2}, both of these evolutions are less dramatic than in the radio band.  This provides an important input to constrain models of jet, accretion disk, and torus emission and their evolution over the history of the universe.  For example, in the basic models of AGN where the spin energy of the black hole is tapped to create the jets \citep[e.g][]{BJ77,BF11}, faster radio evolution than optical would indicate that the spin parameters of black holes were higher in the past since radio emission overwhelmingly results from jets.  Since mid-infrared emission in AGN is some combination of emission from the dusty tori, the jets, and the host galaxies, the significantly less rapid evolution of infrared emission in comparison with radio would confirm that jet emission is a sub-dominant source of infrared emission.  

\subsection{Mid-infrared Luminosity Function}

We also show in \S \ref{localinf} that the local 22 $\mu$m mid-infrared luminosity function of quasars $\psi_{\rm IR}\!(L_{\rm IR}')$ shows a dramatic flattening at luminosities below $10^{31}$\,erg\, s$^{-1}$\,Hz$^{-1}$.  A flattening of this sort is also seen in the 15$\mu$m luminosity function of type-1 and type-2 AGN by \citet{Matute06}, and at 24 $\mu$m by \citet{Xu01}.  Such dramatic flattening of the luminosity function is not seen in optical wavebands at the luminosities probed by this analysis.  However a similar flattening {\it is} seen in the local radio luminosity function at 1.4 GHz luminosities below $10^{30}$\,erg\, s$^{-1}$\,Hz$^{-1}$ by \citet{Kimball11}.  Do these mid-infrared and radio luminosity function flattenings result from the same process?  A simple scaling of the radio emission to the mid-infrared with a synchrotron-like power law spectral index of $\varepsilon=0.5$ would put the mid-infrared luminosity equivalent to the 1.4 GHz break at $\sim10^{28}$\,erg\, s$^{-1}$\,Hz$^{-1}$.  As this is not where the infrared break is observed, the two flattenings seemingly have different physical causes.  This again points to jet emission being a sub-dominant component of the infrared emission.  Rather, it is some phenomenon of the tori or host galaxies that causes the relative scarcity of quasars with mid-infrared luminosities below $10^{31}$\,erg\, s$^{-1}$\,Hz$^{-1}$.  Given that that a differential luminosity function which is flat at the faint end corresponds to a cumulative luminosity function which has a power law slope of -1 at the faint end, we can conclude that the contribution of quasars to the integrated mid-infrared light output in the universe peaks at 22 $\mu$m luminosities of around $2 \times 10^{31}$\,erg\, s$^{-1}$\,Hz$^{-1}$. 

Could the inferred flattening of $\psi_{\rm IR}\!(L_{\rm IR}')$ result from selection effects, in particular the flux limit for optically-identified quasars in the data set used in this analysis?  In principle the analysis of this work accesses the true intrinsic distributions of local luminosities (among other quantities) and corrects for the effects of survey truncations.  It is always possible that the population differs significantly {\it intrinsically} at combinations of $L_{\rm opt}'$, $L_{\rm IR}'$, and $z$ outside of those present in the data set from how it is at combinations represented in the data set.  However, the data set spans two decades in local optical luminosity and local infrared luminosity with significant spread in the $L_{\rm opt}'$, $L_{\rm IR}'$ plane.  We believe it is most likely that the flattening in $L_{\rm IR}'$ is intrinsic in the population of quasars, although cannot rule out from this analysis alone an additional population of low infrared luminosity objects which almost universally have a low optical luminosity. 

\subsection{Integrated Emission from Quasars and Contribution to the Infrared Background Light}

Given the distributions calculated here, we can calculate the total integrated output of quasars in the unverse at 22 $\mu$m.  One should integrate the overall mid-infrared luminosity function $\Psi_{\rm IR}\!(L_{\rm IR},z)$ times the flux corresponding to each luminosity and redshift over all redshifts and luminosities:
\begin{equation}
\mathcal{I}_{\rm 22 \mu m : quasars}=\int_{z} \, dz \, \int_{L_{\rm IR}} \, dL_{\rm IR} { {L_{\rm IR}}  \over {4 \pi {D_L}^2 K_{\rm IR}(z) }  } \, \Psi_{\rm IR}\!(L_{\rm IR},z) .
\end{equation}
In terms of the density evolution function $\rho(z)$ and the mid-infrared luminosity portion $\psi_{\rm IR}\!(L_{\rm IR}/g_{\rm IR}\!(z))/g_{\rm IR}\!(z)$ this is
\begin{eqnarray}
\mathcal{I}_{\rm 22 \mu m : quasars}=\int_{z} \, dz \, \int_{L_{\rm IR}} \, dL_{\rm IR} \, { {L_{\rm IR}}  \over {4 \pi {D_L}^2 K_{\rm IR}(z) }  }
\\
\nonumber  \rho(z) \, { {dV} \over {dz}} \, { {\psi_{\rm  inf} (L_{\rm IR}/g_{\rm IR}\!(z))} \over {g_{\rm IR}(z)} }
\end{eqnarray}
In terms of the local mid-infrared luminosity function $\psi_{\rm IR}\!(L_{\rm IR}')$ we would need the value of this function for the local luminosity corresponding to each luminosity and redshift combination:
\begin{eqnarray}
\mathcal{I}_{\rm 22 \mu m : quasars}=\int_{z} \, dz \, \int_{L_{\rm IR}} \, dL_{\rm IR} \, { {L_{\rm IR}}  \over {4 \pi {D_L}^2 K_{\rm IR}(z) g_{\rm IR}\!(z) }  }
\\
\nonumber  \rho(z) \, { {dV} \over {dz}} \,  \psi_{\rm  inf} (L_{\rm IR}') \left( L_{\rm IR} ,z  \right)
\\
\nonumber {\rm where} \, \, \,  L_{\rm IR}' = { {L_{\rm IR}} \over {g_{\rm IR}\!(z)} }
\label{bgndcont2}
\end{eqnarray}
Carrying out this integration results in a calculated value of $\mathcal{I}_{\rm 22 \mu m : quasars} = 4.4 \pm 2.8 \times 10^{-24}$ W m$^{-2}$ sr$^{-1}$ Hz$^{-1}$.  In $\nu \mathcal{I}_{\nu}$ units this is $\nu \mathcal{I}_{\rm 22 \mu m : quasars} = 6.0 \pm 3.8 \times 10^{-11}$ W m$^{-2}$ sr$^{-1}$, which can be compared to e.g. the results obtained by \citet{Matute06} at 15 $\mu$m who report a value of (4.2 --- 12.1) $\times 10^{-11}$ W m$^{-2}$ sr$^{-1}$ for type-1 AGN and (5.5 --- 14.6) $\times 10^{-11}$ W m$^{-2}$ sr$^{-1}$ for type-2 AGN.  It would be enlightening to compare the number obtained here for the integrated output of quasars at 22 $\mu$m to the total cosmic infrared background light (CIB) level at this wavelength.  Unfortunately a gap exists in reported measurement of the CIB between 3.5 $\mu$m and 60 $\mu$m, with reported values of $11 \pm 3.3 \times 10^{-9}$ W m$^{-2}$ sr$^{-1}$ at the former \citep{Gorjian00} and $28.1 \pm 8.8 \times 10^{-9}$ W m$^{-2}$ sr$^{-1}$ at the latter \citep{Fink00}. Taking an intermediate value between these two as an estimate of the CIB at 22 $\mu$m would indicate that quasars contribute less than one percent of the total integrated mid-infrared light output in the universe.  Alternately, one could also compare the value obtained here for the output of quasars to the total output calculated from source counts for all sources at this wavelength.  \citet{Papovich04} calculate this output from all sources to be $2.7 (+1.1 / -0.7) \times 10^{-9}$ W m$^{-2}$ sr$^{-1}$ at 24 $\mu$m, which, ignoring any spectral shape between 24 $\mu$m and 22 $\mu$m and given the value obtained here, would make quasars responsible for between 1\% and 5\% of the total output from 22 $\mu$m sources.

\subsection{Luminosity Correlations}

As discussed in \S \ref{roevsec}, the mid-infrared and optical luminosities are highly correlated, but the power-law correlation index between the mid-infrared and optical luminosities ($\sim$0.8) is less than that found previously for the radio and optical luminosities ($\sim$1.0).  As discussed in \S \ref{ifcorr}, the subject of luminosity-luminosity correlations is complicated and it is not straightforward to determine how much of these correlations are intrinsic to the waveband emissions in the population and how much are induced by similar redshift evolutions and the truncations of the data set.  We will explore this issue in a future work.  For the present, we can speculate that {\it if} these correlations are intrinsic (or if the induced portion is roughly the same in the infrared-optical case as the radio-optical case), the radio-optical correlation being more powerful than the infrared-optical correlation could support the idea that the mass and/or spin of the black hole affect the size and/or temperature of the accretion disk and power of the jets more than they affect the size and/or temperature of the torus.  A full understanding of the true nature of luminosity-luminosity correlations in AGN, and an extension of these considerations to the X-ray band, will be useful in exploring these and other questions.

\acknowledgments

Funding for the SDSS and SDSS-II has been provided by the Alfred P. Sloan Foundation, the Participating Institutions, the National Science Foundation, the U.S. Department of Energy, the National Aeronautics and Space Administration, the Japanese Monbukagakusho, the Max Planck Society, and the Higher Education Funding Council for England. The SDSS Web Site is http://www.sdss.org/.  This publication makes use of data products from the Wide-field Infrared Survey Explorer, which is a joint project of the University of California, Los Angeles, and the Jet Propulsion Laboratory/California Institute of Technology, and {\it NEOWISE}, which is a project of the Jet Propulsion Laboratory/California Institute of Technology. {\it WISE} and {\it NEOWISE} are funded by the National Aeronautics and Space Administration.  We thank the referee for very insightful comments.

\end{document}